\documentclass[manuscript]{acmart}
\usepackage{url}
\usepackage{graphicx}
\usepackage{algorithm,algorithmic}
\usepackage{cancel}
\usepackage{subfigure}
\usepackage{multicol}
\newtheorem{definition}{Definition}

\usepackage[font=small,labelfont=bf]{caption}

\usepackage{comment}
\AtBeginDocument{%
  }

\begin{document}

\title{Empowering Volunteer Crowdsourcing Services: A Serverless-assisted, Skill and Willingness Aware Task Assignment Approach for Amicable Volunteer Involvement}

\author{Riya Samanta, Biswajeet Sethi, Soumya K. Ghosh }
\affiliation{%
  \institution{Indian Institute of Technology Kharagpur \country{India}}}
\email{riya.samanta@iitkgp.ac.in, biswajeet.sethi@iitkgp.ac.in, skg@cse.iitkgp.ac.in}


\renewcommand{\shortauthors}{Riya-2 et al.}

\begin{abstract}
Volunteer crowdsourcing (VCS) leverages citizen interaction to address challenges by utilizing individuals’ knowledge and skills. Complex social tasks often require collaboration among volunteers with diverse skill sets, and their willingness to engage is crucial. Hence, matching tasks with the most suitable volunteers remains a significant challenge. VCS platforms face unpredictable platform demands in terms of tasks and volunteers requests, complicating resource requirement prediction for volunteer to task assignment process. To address these challenges, we introduce the Skill and Willingness-Aware Volunteer Matching (SWAM) algorithm, which allocates volunteers to tasks based on skills, willingness, and task requirements and developed a serverless framework to deploy SWAM. Our method outperforms conventional solutions, achieving a 71\% improvement in end-to-end latency efficiency. We achieved a 92\% task completion ratio and reduced task waiting time by 56\%, with an overall utility gain 30\% higher than state-of-the-art baseline methods. This framework will contribute to generating amicable volunteer and task matches, supporting grassroots community coordination and fostering citizen involvement, ultimately contributing to social good.
\end{abstract}

\ccsdesc[500]{Human-centered computing~Collaborative and social computing theory, concepts and paradigms}
\ccsdesc[500]{Information systems~Crowdsourcing}
\ccsdesc[500]{Cloud computing}

\keywords{Volunteer Crowdsourcing, Skill-oriented, Willingness, Task Assignment, Serverless Computing }


\maketitle

\section{Introduction}
\vspace{-0.05in}
Crowdsourcing has become a powerful catalyst for social change and tackling complex societal issues. By leveraging the collective intelligence and resources of a diverse group of people, crowdsourcing initiatives can create innovative solutions to challenges such as situational awareness, environmental sustainability, and public health. 
Crowdsourcing enables global collaboration in many ways, such as using collective knowledge to set sustainable goals, involving citizens in monitoring biodiversity, or mobilizing volunteers to map crisis areas for humanitarian aid. Moreover, it empowers communities to actively participate in decision-making processes as in democrat \cite{tumedei2021promoting}.
 
Serverless computing, a cloud computing execution model, with its on-demand scalability and pay-per-use protocol, has immense potential to contribute to social good in various ways. By eliminating the need for provisioning and managing servers, it reduces infrastructure costs, making it more accessible for nonprofits and social enterprises with limited resources to develop and deploy applications aimed at solving societal challenges \cite{sethi2023lcs}. Serverless architecture enables rapid prototyping and deployment, facilitating the creation of innovative solutions for humanitarian aid, disaster response, healthcare, education, and environmental conservation. Its event-driven nature also allows for efficient processing of real-time data, enabling organizations to respond swiftly to emerging crises and deliver critical services to communities in need. Overall, serverless computing empowers organizations to concentrate on their missions without being burdened by infrastructure constraint. This ultimately has a positive impact on society.

In this work, we focus on volunteer crowdsourcing (VCS) \cite{Sama2212:Volunteer} as a pivotal aspect of improving social impact. Volunteer crowdsourcing harnesses citizens' voluntary interaction to tackle challenges by leveraging individuals' knowledge and skills. The integration of smart mobile devices with Web 2.0 technology has significantly enhanced the effectiveness of \textit{skill-based volunteer crowdsourcing} in domains such as healthcare, emergencies, sustainable development, situational awareness, and social awareness. In Asia, several Non-governmental organizations (NGOs) including Goonj, ActionAid, and Smile \cite{goonj,actionaid,smile} have successfully utilized volunteer crowdsourcing to tap into the expertise and skills of volunteers. Globally, platforms like VolunteerMatch and Idealist \cite{volunteermatch,idealistic} facilitate the alignment of volunteers' interests and skills with the needs of various NGOs. However, matching tasks with the most suitable volunteers remains a significant challenge. Complex tasks often necessitate collaboration among volunteers with diverse skill sets \cite{Song2020}. In addition to skills, the willingness of volunteers to engage in community activities is crucial \cite{samanta2021swill}. Despite their competence, a volunteer's reluctance can hinder success, impacting both task outcomes and the overall utility of the VCS platform. This issue is particularly critical for volunteer services operating within limited budgets and requiring prompt, efficient action. Volunteers may also face costs for transportation, accommodation, sustenance, healthcare, and equipment \cite{Sama2212:Volunteer}, highlighting the importance of considering volunteer remuneration. These VCS platforms regularly experience a dynamic influx of tasks and volunteers. Determining the required computing resources for volunteer to task matching process is challenging due to unpredictable platform demands in terms of volunteer and tasks requests \cite{miao2022dynamically}. Therefore, these platforms must strategically allocate tasks to volunteers in a scalable manner. Small businesses, start-ups, and NGOs are actively seeking cost-effective and quickly deployable solutions to enhance their operational efficiency.

\begin{figure}[!t]
 \centering
  \includegraphics[width=0.9\columnwidth]{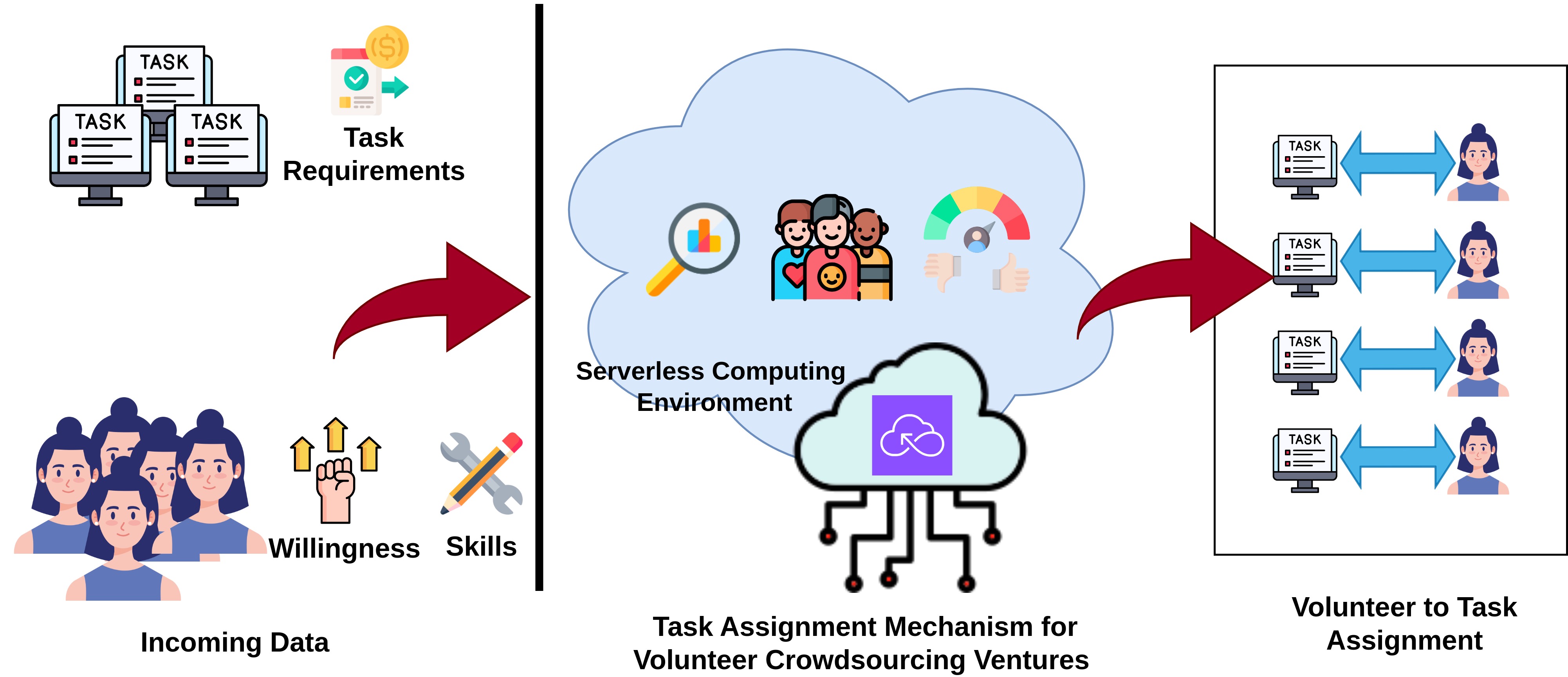} 
  \caption{Overview of proposed Serverless-assisted Framework for task assignment in Volunteer Crowdsourcing Paradigm}
  \label{overview}
  \vspace{-0.2in}
\end{figure}

In this study, we propose a serverless-assisted framework for task assignment in the VCS paradigm. This framework leverages the benefits of serverless computing to address the challenges of managing dynamic task assignments for a crowdsourcing system. The \textbf{Figure. \ref{overview}} shows the overview of the proposed framework.

\subsection{Existing Limitations and Research Gap}
The academic research community has largely overlooked the potential of integrating serverless architecture with crowdsourcing applications, despite its significant potential to advance social good (see Related Work, Section \ref{relatedwork}). Existing literature on skill-based task assignments often neglects critical factors such as skill diversity and workers' willingness, which are crucial for effective task assignment policies in volunteer crowdsourcing (VCS) ventures. These factors become particularly critical in online (dynamic) assignment modes, reflecting real-world scenarios where participants exhibit dynamic arrival and departure patterns following probability distributions. In such online modes, achieving dynamic resource scalability with minimal human intervention is essential. However, current literature lacks discussions on the scalability of crowdsourcing platforms from the deployment standpoint of task assignment mechanisms. Integrating serverless architecture with crowdsourcing applications offers promising solutions to these challenges, enabling platforms to dynamically scale and efficiently handle fluctuating workloads of task assignment decisions in real-world applications.


\subsection{Contributions:} 
In the pursuit of addressing the above mentioned limitations and challenges in the domain of skill-oriented task assignment paradigm for VCS ventures, the main contributions of this paper are: \\ 
\noindent(1) We propose the \textit{Skill and Willingness-Aware Volunteer Matching (SWAM)} algorithm, which allocates volunteers to tasks based on their skills, willingness, and tasks' skill requirements. 

\noindent(2) We develop a \textit{serverless} framework to deploy the SWAM algorithm , enhancing end-to-end performance by reducing latency and average task waiting time.

\noindent(3) We demonstrate the effectiveness of using a serverless ecosystem for executing SWAM, focusing on latency and task waiting time. We compare SWAM with state-of-the-art skill-oriented task assignment algorithms, evaluating task completion ratio and overall platform utility using a real dataset from Meetup \cite{meetup}.

The remainder of this paper is structured as follows: Section \ref{pf} presents preliminaries required for designing the proposed framework. In Section \ref{sol}, gives the details of the methodology including the working of proposed SWAM algorithm and outline of the serverless deployment framework that enables our SWAM algorithm to be deployed. Section \ref{SL} includes an evaluation of the performance of the proposed model compared to the baseline algorithms. In Section \ref{relatedwork}, we review the relevant literature and prior work in the field. Finally, in Section \ref{conc}, we provide the conclusion and future work.

\section{Preliminaries}\label{pf}
We proceed to present necessary definitions, preliminaries and important data structures needed in this study. 

\begin{definition}
(\textbf{Volunteer}) The set of interested volunteers is represented by $V=\{v_{1},v_{2},v_{3},....,v_{n}\}$, where each volunteer $v_{j}$ is represented by a tuple $<q_{v},p_{v},\delta_{v},e_{v}>$. The tuple includes the volunteer's skill set ($q_{v}$), desired remuneration ($p_{v}$), arrival time ($\delta_{v}$), and expected time-to-live ($e_{v}$).
\end{definition}

\begin{definition}
(\textbf{Task}) The set of tasks is represented by $T=\{t_{1},t_{2},t_{3},....,t_{m}\}$, where each task $t_{i}$ is represented by a tuple $<q_{t},b_{t},\delta_{t},e_{t}>$. The tuple includes the required skill set ($q_{t}$), pre-defined budget ($b_{t}$), arrival time ($\delta_{t}$), and expected time-to-live ($e_{t}$). The task is either withdrawn from the system or re-posted after $e_{t}$ expires.
\end{definition}

The proposed framework operates under the assumption that a task's completion depends on the coverage of skills required for that task. Drawing from the research by \cite{Song2020,samanta2021swill, Sama2212:Volunteer,Cheng2016}, it is assumed that both the skills required for a given task and the skills possessed by a volunteer are consistently drawn from a predefined and constant set of skills, denoted as $Q$. This implies that for all volunteers $v \in V$ and tasks $t \in T$, the sets of skills $q_{v}$ and $q_{t}$ are subsets of $Q$. Additionally, following the concept of one-to-many mapping as discussed in \cite{Song2020,samanta2021swill, Sama2212:Volunteer,Cheng2016}, it is possible for a task to have multiple volunteers at a given time frame, but the reverse scenario is not true

\begin{definition}
(\textbf{Willingness}) It is a crucial aspect in determining the participation of a volunteer in a task. For a volunteer, the willingness to participate in a task denoted as $W(v,t)$, is defined as the probability of a volunteer's intention to participate towards the completion of a task. The willingness is formulated as:
\begin{equation}\label{eq1}
W(v,t) = \frac{1}{1 + \exp(-\omega(v,t))}
\end{equation}
where $\omega(v,t)$ is the summation of the volunteer's efficiency ($eff_{v}$) and bias ($\beta_{v}$) towards the task, given by:
$\omega(v,t) = eff_{v} + \beta_{v}$
\end{definition}

The willingness is measured by the combination of two factors: the volunteer's \textit{efficiency} ($eff_v$) and \textit{bias} ($\beta_v$). Assuming that the variables $\beta_v$ and $eff_v$ range from 0 to 1, $\omega(v,t)$ falls within the range of 0 to 2, ensuring that $W(v,t)$ always ranges between 0.5 and 1. The efficiency ($eff_v$) represents the volunteer's proficiency in the required skills for the task, while the bias ($\beta_v$) expresses the volunteer's preference for the task. Efficiency can be derived from the volunteer's job completion history or performance ratings, and bias can be inferred from the types of tasks the volunteer has previously accepted or completed. For this study, we assume that these data are available to the algorithm from the utilized dataset. The willingness of a volunteer is task-specific and varies based on factors influencing the volunteer's decision-making, such as payment, duration, and prior experience. A highly willing volunteer is more likely to dedicate time and effort to tasks, resulting in improved outcomes, which is crucial for enhancing social community services \cite{samanta2021swill}.

\begin{definition}
(\textbf{Utility Score}) It provides a numerical representation of the suitability of a task for a volunteer, with higher scores indicating a better fit. Given a volunteer $v$ and a task $t$, the utility score of $v$, denoted by $U_{v,t}$ is calculated by the formula as follows:
\begin{equation}\label{util_v}
   U_{v,t} =  b_{t} \times W(v,t)
\end{equation}
\end{definition}
It is obvious that, from a volunteer standpoint, a task with a higher fund limit can be expected to cover the incurred expenditure to a certain threshold or to pay $v$'s share of remuneration without fail. Thus, $b_{t}$ is positively correlated with $U_{v,t}$. Additionally, the willingness factor has a positive impact on the $U_{v,t}$ \cite{samanta2021swill}.

\begin{definition}
(\textbf{Batch-based Assignment}) Given a set of fixed rotation intervals, Time = $[X_{0},X_{1},X_{2},...X_{k}]$, where $X_{k} = k \times X_{1}$, a batch-based assignment involves an assignment map $\mu$ represented as a bipartite graph $(V, T, E)$, where $V=\{v_{1},v_{2},v_{3},....,v_{n}\}$ denotes volunteer nodes and $T=\{t_{1},t_{2},t_{3},....,t_{m}\}$ denotes task nodes. The edges in the bipartite graph $E$ are defined as pairs $(v, t) \in V \times T$.
\end{definition}

The batch-based strategy combines elements of both \textit{offline} and \textit{online} methods, where tasks and volunteers accumulate over a specified time interval before assignment decisions are made. This periodic process aims to minimize assignment costs while maintaining responsiveness in real-time scenarios \cite{sethi2023scalable}.

\begin{definition}
(\textbf{Skill-Task Mapper}) is defined as a bipartite graph denoted by $G_{ST}=(Q, T, E)$, where an edge $(x,y) \in E$ exists if and only if a skill $x \in Q$ is required for a task $y \in T$.
\end{definition}
The primary intent of the graph is to serve as an index for the skills required for each task and to monitor the extent to which skill requirements are met. The Skill-Task Mapper is implemented as a two-dimensional matrix that undergoes dynamic updates subsequent to each allocation. Only the final row  encompasses the budgetary value of every task in order. The process of constructing the matrix involves amalgamating all active tasks to form the column, while each tuple is represented by every skill from the set $Q$. Consequently, the dimensions of the $G_{ST}$ matrix will be $(|Q|+1) \times |T|$. 

\begin{definition}
(\textbf{Volunteer-Skill Mapper}) is a  bipartite graph represented by $G_{VS}=(Q^{'}, T, E)$, where an edge $(x,y)$ exists if and only if a skill $x \in Q^{'}$ is required for any task $y \in T$. $G_{VS}$ is constructed in specific for a volunteer under observation. Thus, $Q^{'}$ is equivalent to $q_{v}$, which is the skill set available with the volunteer $v$. In the worst case, $Q^{'} \equiv Q$.
\end{definition}
This  data structure is predominantly utilised for identifying tasks that are most compatible with the skill set possessed by a volunteer. The implementation of $G_{VS}$ employs a two-dimensional matrix to map a volunteer's available skills to the required skill set of available tasks. The $G_{VS}$ will have dimensions of $|q_{v}| \times |T|$. The production of $G_{VS}$ can be derived in a direct manner from $G_{ST}$ as a sub-graph.

\begin{definition}
(\textbf{Volunteer Utility Vector}) is a  vector that stores volunteer utilities for each task. It is denoted by $UtilVec$. The utility score is calculated using Equation-\ref{util_v}.
\end{definition}
Thus the corresponding $UtilVec$ of $v_{3}$ is given by:
\begin{equation}\label{utilvec}
\begin{gathered}
UtilVec=[U_{v,t_{1}},U_{v,t_{2}},...,U_{v,t_{n}}]\\
=[(b_{t_{1}} \times W(v,t_{1})), (b_{t_{2}} \times W(v,t_{2}))...(b_{t_{n}} \times W(v,t_{n}))] 
\end{gathered}
\end{equation}
$UtilVec$ is used in the SWAM's allocation decision. and its length is equal to $|T|$.

\section{Methodology}\label{sol}
Our work employed the Server Allocated Task (SAT) model \cite{Cheng2016} to effectively match volunteers to tasks. We used \textit{AWS Lambda} \cite{AWSLambda} for serverless computation and \textit{AWS S3} \cite{S3} for storing incoming, intermediate, and result data. Given the continuous arrival of task and volunteer data, we implemented a \emph{batch-based split and allocate strategy}. At the start of each time interval ($X_i$), based on data received in the previous batch cycle ($X_{i-1}$), Lambda functions split the incoming data into chunks. We used \textit{Amazon EventBridge} \cite{EB} to trigger these functions periodically.

The splitting process could be regulated by predetermined rules or occur stochastically. In our approach, volunteers were partitioned into $\kappa$ subgroups, as the number of volunteers typically exceeds the number of tasks on most crowdsourcing platforms, including the Meetup dataset used in this study. The entire task data is processed to construct the bipartite graph $G_{ST}$ (see Definition 7). This graph is then replicated and distributed to each of the $\kappa$ Lambda functions, denoted as $L_{SWAM}$, along with one of the $\kappa$ partitioned volunteer datasets. The $\kappa$ allocation maps are generated by concurrently executing $\kappa$ instances of the SWAM algorithm as $\kappa$ Lambda functions that is $L_{SWAM}$. Finally, to consolidate these multiple allocation maps into a single final result, we execute a Lambda function, denoted as $L_{Consol}$.

\subsection{Skill and Willingness-Aware Volunteer Matching Algorithm} 
This section describes the working of the SWAM algorithm (see Algorithm \ref{algo1}) in detail with example settings depicted in Table \ref{tasks} and Table \ref{vols}. The SWAM algorithm takes as its input the pre-processed $G_{ST}$ and a partitioned volunteer data denoted as $V^{'}$, where $V^{'} \in V$. The resulting output comprises a penultimate allocation map denoted as $\mu^{'}$ and an updated $G^{'}_{ST}$.  The map $\mu$ essentially functions as a dictionary that maps tasks to the volunteers assigned to them.

\begin{table}[h]
\centering
\caption{Details of the tasks}
\vspace{-0.1in}
\begin{tabular}{p{2.5cm}p{4cm}p{2cm}}
\hline
\textbf{Tasks} & \textbf{Skill requirements} & \textbf{Budget} \\
\hline
$t_{1}$ & Teaching, Nutritionist, Nursing & \$650 \\
$t_{2}$ & Child care, Teaching, Cooking & \$500 \\
\hline
\end{tabular}
\label{tasks}
\vspace{-0.1in}
\end{table}

\begin{table}[h]
\centering
\caption{Details of the volunteers}\label{vols}
\vspace{-0.1in}
\begin{tabular}{p{2.5cm}p{4cm}p{2cm}}
\hline
{\bf Volunteers} & {\bf Skills} & {\bf Remuneration} \\
\hline
$v_{1}$         & Child-care, Guitarist, Nursing       & \$25   \\
$v_{2}$         & Cooking, Nursing, Nutritionist  & \$20 \\
$v_{3}$         & Teaching, Cooking, Nursing   & \$30         \\
\hline
\end{tabular}
\vspace{-0.1in}
\end{table}

The initial stage involves generating a localised replica of $G_{ST}$ to preserve the integrity of the original $G_{ST}$ throughout the algorithmic process, as it is required for subsequent stages. The volunteer set is traversed, the one with the lowest remuneration is picked, and his or her $G_{VS}$ is constructed. The appropriate volunteer from Table \ref{vols} is $v_{2}$ (seeking a minimum pay of \$20). Next, the $UtilVec$ of $c$ is created by the equation-\ref{utilvec}. We assume that all volunteers have a $W(c,t)$ of $0.5$. The $UtilVec$ of $v_{2}$ for tasks $t_{1}$ and $t_{2}$ is $[(650*0.5),(500*0.5)]$ =$[325,250]$. The column-wise sum of $G_{VS}$ matrix is performed. Thus, the resultant $col\_sum$ for $G_{VS}$ of $v_{2}$ is $[2,1]$. To obtain the vector $t_{pref}$, each $i^{th}$ element of the $col\_sum$ vector is multiplied by each $i^{th}$ element of the $UtilVec$ vector. As a result, $t_{pref}$ for $v_{2}$ is $[(2 \times 325), (1 \times 250)]$ = $[650,250]$. If $col\_sum$ of $v_{2}$ had been $[2,3]$, $t_{pref}$ of $v_{2}$ would have been $[650,750]$. The while loop begins. Initially, the argument of the element with the highest value is chosen from $t_{pref}$. The argument, in this case, is $t_{1}$. The current budget of $t_{1}$ allows for volunteer $v_{2}$. As a result, $v_{2}$ is assigned to $t_{1}$, and $b_{t_{1}}$ is updated. A new entry for the pair $(v_{2}, t_{1})$ is also added to the $\mu^{'}$. Then, the control comes out of the while loop. If $b_{t_{1}}$ was insufficient for recruiting $v_{2}$, $v_{2}$ is removed from $t_{pref}$ and the next immediate argument (a.k.a task) is chosen from $t_{pref}$. This continues until $t_{pref}$ is empty or completely traversed. At the end, $G^{'}_{ST}$ is updated.  Finally, SWAM concludes by returning $G^{'}_{ST}$ and $\mu^{'}$.

\begin{algorithm} [h]
 \caption{SWAM (Skill and Willingness-Aware Volunteer Matching)} \label{algo1}
 {\algsetup{linenosize=\tiny}
  \scriptsize{
 \begin{algorithmic}[1]
 \renewcommand{\algorithmicrequire}{\textbf{Input:}}
 \renewcommand{\algorithmicensure}{\textbf{Output:}}
 \REQUIRE Volunteer $V^{'}$, Skill-Task Mapper $G_{ST}$
 \ENSURE Allocation Map $\mu^{'}$, $G_{ST}$
 \\
\STATE Start
\STATE $G^{'}_{ST}$ $\gets$ $G_{ST}$
\FORALL{ $c \in V^{'}$}
\STATE Select volunteer $c$ from $V^{'}$ with $min(p)$
\STATE Generate $G_{VS}$ of $c$ 
\STATE Generate the $UtilVec$ vector (Equation-\ref{utilvec})
\STATE $col\_sum$ $\gets$ Column-wise sum of $G_{VS}$
\STATE $t_{pref} \gets$ Null
\FOR{ $i=0$ to $|T|$}
\STATE $t_{pref}[i]$ $\gets$ $col\_sum[i] \times UtilVec[i]$
\ENDFOR
\WHILE{$t_{pref}$ is not empty}
\STATE $t_{reco}$ $\gets$ $argmax(t_{pref})$
\IF {$b_{t_{reco}}$ $\geq$ $p_{c}$}
\STATE $t_{a}$ $\gets$ $t_{reco}$ 
\STATE Add allocation $(c,t_{a})$ to $\mu^{'}$
\STATE $b_{t_{a}} = b_{t_{a}} - p_{c}$
\STATE break
\ELSE 
\STATE Remove $t_{reco}$ from list $t_{pref}$
\ENDIF
\ENDWHILE
\STATE Update $G^{'}_{ST}$
\ENDFOR
\RETURN $G^{'}_{ST}$, $\mu^{'}$
\STATE End
\end{algorithmic}
}}
\end{algorithm}
\vspace{-0.1in}

\subsection{Serverless Backbone} \label{serverless}
Our framework integrates crowdsourcing with serverless computing platforms, particularly leveraging the \textit{Function-as-a-Service} (FaaS) model known for its automatic scalability and resource provisioning capabilities. The modularity of FaaS architecture contrasts with traditional monolithic applications, making it suitable for managing dynamic data arrival patterns in real-time. This capability is crucial for efficiently handling sudden data influxes without manual intervention, a key advantage of serverless computing over conventional cloud platforms where scalability configuration is developer-dependent. Our serverless deployment workflow involves several key steps:
(i) Task and volunteer data are initially stored as separate objects in \textit{AWS S3} buckets.
(ii) Lambda functions ($L_{t}$ for tasks and $L_{v}$ for volunteers) process these data to generate the bipartite graph $G_{ST}$ and segmented volunteer data chunks $V_{split}$ respectively. The number of segments for volunteer data can be adjusted based on dataset size and complexity.
(iii) Each segment of volunteer data is stored in \textit{AWS S}3, with each bucket mapped to a dedicated Lambda function ($L_{SWAM}$). Concurrently, $G_{ST}$ is stored in a separate \textit{S3} bucket.
(iv) The arrival of segmented volunteer data in \textit{S3} triggers respective $L_{SWAM}$ functions, which fetch $G_{ST}$ to produce assignment maps $\mu^{'}$ and update skill-task mappings $G^{'}_{ST}$.
(v) These processed outputs are passed directly to $L_{Console}$ without intermediate \textit{S3} storage, ensuring efficient data flow.
(vi) Finally, $L_{Console}$ consolidates data, removes redundancy, and generates the final assignment map $\mu$.

The orchestration of this serverless workflow is managed by \textit{AWS Step Functions}, providing visual workflows for coordinating distributed application tasks.

\begin{figure}[!t]
 \centering
  \includegraphics[width=0.9\textwidth]{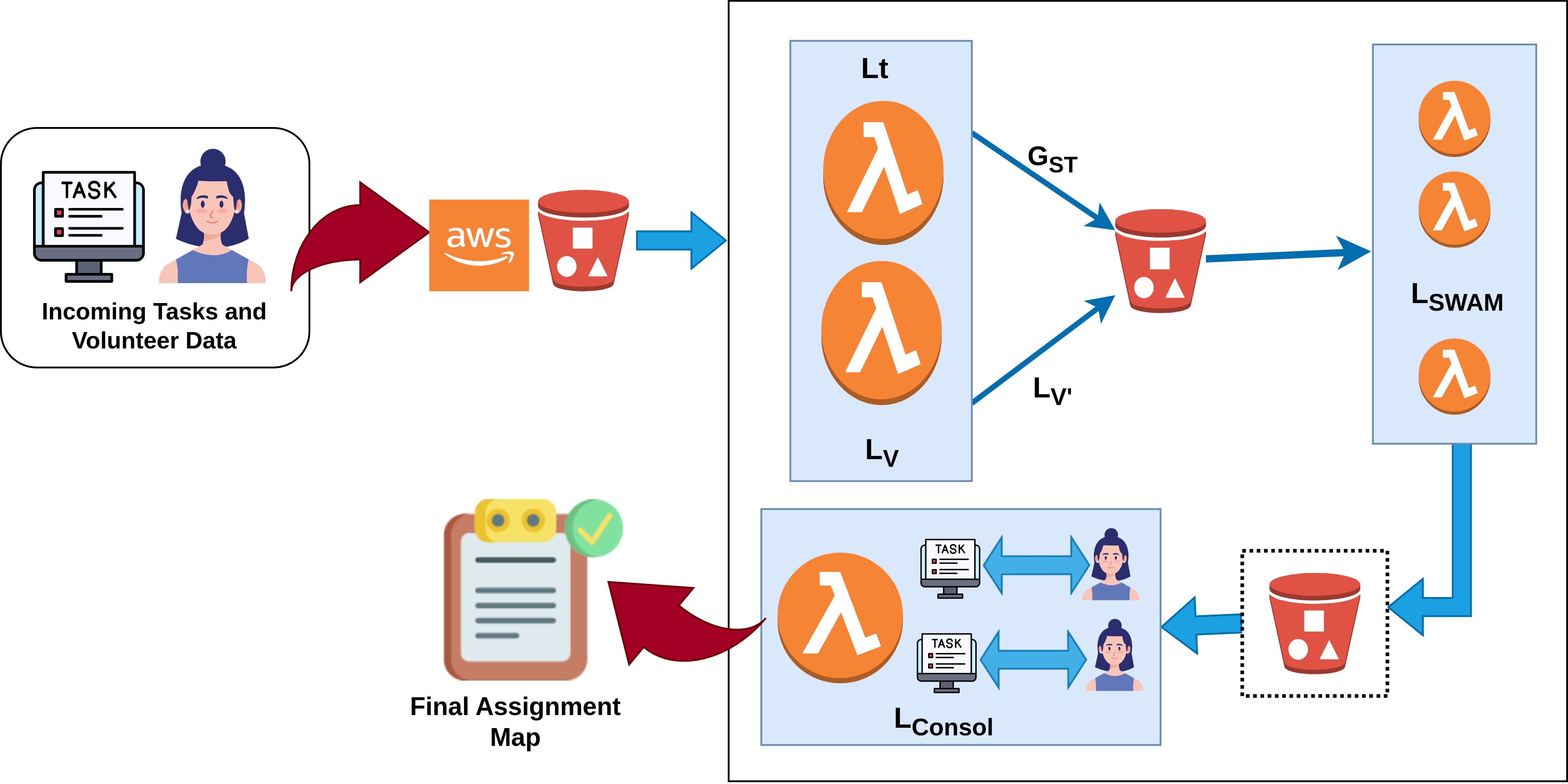} 
  \vspace{-0.1in}
  \caption{Serverless Deployment Framework for Task Assignment Mechanism }
  \label{framework}
   \vspace{-0.1in}
\end{figure}

\vspace{-0.1in}
\section{Performance Evaluation}\label{SL}
\vspace{-0.05in}
\subsection{Dataset}
\vspace{-0.05in}
We use the benchmark event-based social network dataset \textbf{Meetup} \citep{meetup} to simulate our problem, which is a common practice in evaluation of skill-oriented crowdsourcing platform \citep{Xie2023,sethi2023scalable}. In this dataset, the events are utilised as tasks and the users are designated as crowd workers. The labels are perceived as skills. We pruned out the skills of the tasks which cannot be covered and that of the crowd workers which are not demanded much by the tasks. There are 1234 tasks ($|T|$) and the total number of skills ($S$) is 554. Each task requires 5 to 10 skills, giving an average of 3.6 skills per task. The mean task budget following Gaussian distribution is \$428, with a standard deviation of \$255. There are 3275 ($|V|$) crowd workers, each having 1 to 5 skills, giving an average of 1.7 skills per crowd worker. The mean remuneration of each crowd worker following Gaussian distribution is \$40, with a standard deviation of \$50. The $eff_{v}$ and $\beta_{v}$ are generated using a uniform distribution between 0 and 1.

\begin{figure}[!t]
 \centering
\includegraphics[width=0.45\textwidth,height=2in]{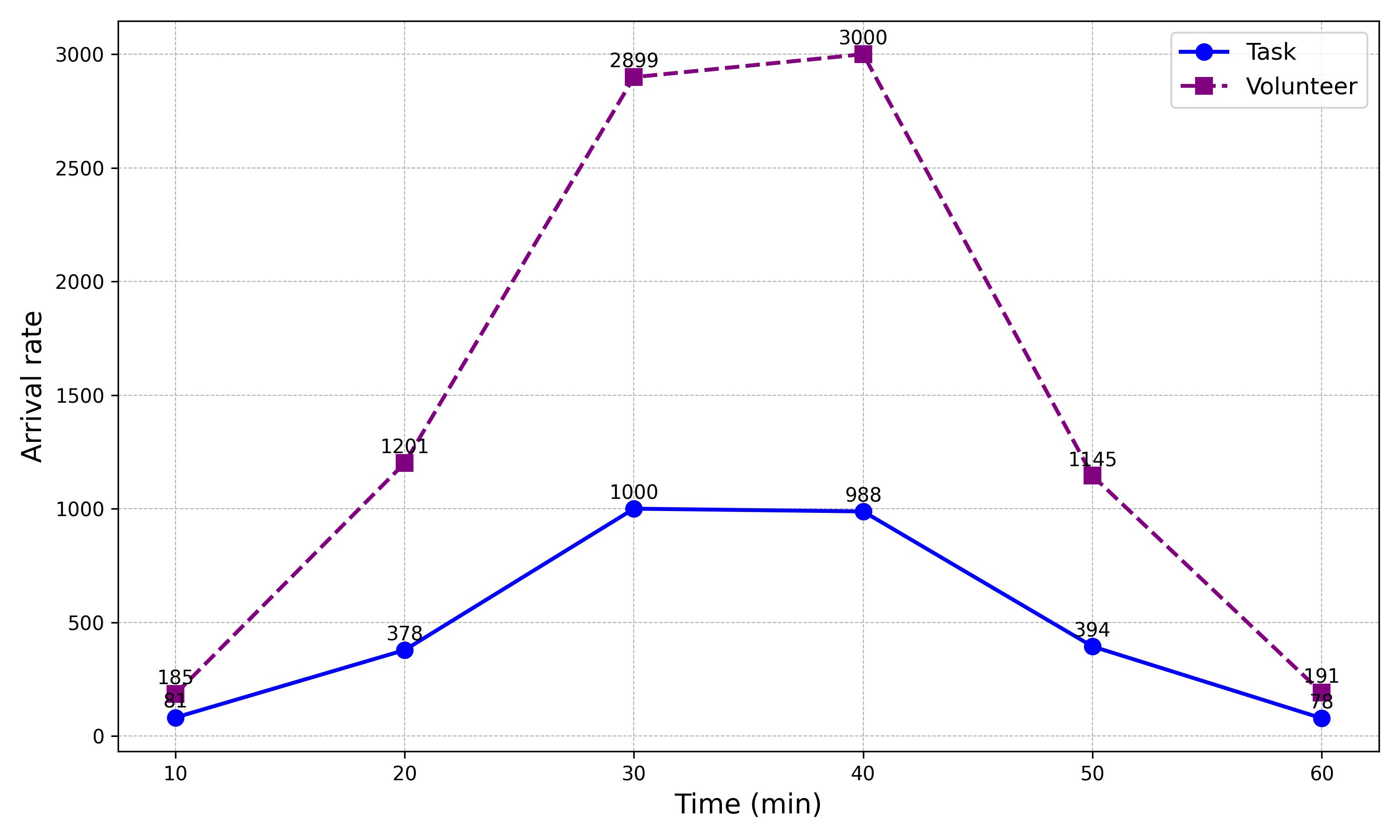}
  \vspace{-0.16in}
  \caption{Arrival pattern of tasks and volunteers}
  \label{arrival_trends}
\vspace{-0.1in}
\end{figure}

\textbf{Figure \ref{arrival_trends}} illustrates the arrival patterns for both tasks and volunteers, using data generated from a Normal Distribution \cite{patel1996handbook} to demonstrate the auto-scaling capability of the proposed framework compared to baseline methods. The x-axis represents time in minutes within an hour interval, while the y-axis shows the arrival rate. The mean is set to 30 minutes, and the standard deviation is 10 minutes. Initially, there is a positive slope for both tasks and volunteers, indicating increasing arrival rates. This upward trend facilitates the evaluation of serverless resource up-scaling. Conversely, a downward slope starting at the 40-minute mark allows for the evaluation of resource down-scaling. The experiment uses a batch size of 10 minutes.

\subsection{Experimental Results and Analysis}\label{results}
The evaluation was conducted in two phases. In the first phase, we compared the performance of the SWAM algorithm locally and in a serverless environment, capturing the \textbf{end-to-end latency} and \textbf{waiting time} in microseconds. In this context, waiting time is defined as the total time a task has to wait before it gets assigned with all the volunteers it needed \cite{sethi2023scalable}. In the second phase, we compared the proposed SWAM algorithm with two baseline methods: \textbf{Online Greedy} (\textbf{OG}) \cite{Song2020} and \textbf{initial Volunteer Task Mapping} (\textbf{i-VTM}) \cite{Sama2212:Volunteer}. The comparison was based on the \textbf{task completion ratio} and \textbf{overall utility score}. The task completion ratio for a specific batch instance is the ratio of successfully completed tasks (those meeting skill requirements) to the total available tasks. A task is considered `complete' only if all associated skills are addressed. The utility score is calculated using Equation \ref{util_v}.

\begin{table}[h]
\centering
\caption{Experimental Settings}
\vspace{-0.1in}
\label{settings}
\begin{tabular}{c p{0.6\linewidth}}
\hline
\textbf{Local Configurations} & Processor - Intel i3 dual-core\\
                              & CPU Frequency - 2GHz\\
                              & RAM - 4GB\\
                              & OS - Windows 10\\
                              & Programming Language - Python 3.9\\
\hline
\textbf{Serverless Configurations} & Platform - Amazon Web Service\\
                                   & Computational Unit - AWS Lambda\\
                                   & Storage unit - AWS S3\\
                                   & Iam Role - S3FullAccess; CloudwatchFullacess\\
                                   & Lambda memory - 128 MB\\
                                   & CPU - Auto provisioned by AWS\\
                                   & Programming Language - Python 3.9\\
\hline
\end{tabular}
\vspace{-0.1in}
\end{table}

In the first phase of our experimentation, we evaluated the SWAM algorithm's performance in a scalable serverless architecture with stateless modular parallel executing functions, comparing it against a local environment. Table \ref{settings} details the characteristics of both deployment environments. We focused on end-to-end latency as the evaluative parameter. \textbf{Figure 4(a)} illustrates the behavior of the SWAM algorithm in both environments. In the local environment, SWAM exhibited a latency ranging from 3665 ms to 4433 ms. In contrast, leveraging serverless functionalities reduced latency to a range of 1022 ms to 1322 ms. \textbf{Figure 4(b)} presents the average waiting times observed for SWAM. In the local environment, average waiting times varied between 786 ms and 988 ms. With the serverless ecosystem, maximum and minimum waiting times were reduced to 487 ms and 401 ms, respectively.

During the second phase of our experimentation, we conducted a comparative analysis among the SWAM algorithm, OG algorithm, and i-VTM algorithm to assess their effectiveness in achieving task completion ratios and overall utility scores.\textbf{ Figure 5(a)} illustrates the comparison of completion ratios. The SWAM algorithm achieved an average completion rate of 0.92, which is significantly higher compared to OG and i-VTM. Specifically, SWAM showed a fourfold increase over OG and a one-and-a-half-fold increase over i-VTM. As depicted in the figure, after the initial 40-minute period, there was a decrease of approximately 60\% in the number of available volunteers, while the number of tasks experienced a 50\% decline. Consequently, fewer volunteers were available to match with tasks based on skill requirements. Additionally, tasks from previous batches accumulated in the pool of incomplete tasks, either due to inadequate funding or unmet skill requirements among participating volunteers. 

\begin{figure}[!t]
\subfigure[End-to-end latency (in ms)]
{\includegraphics[width=0.45\textwidth,height=2in]{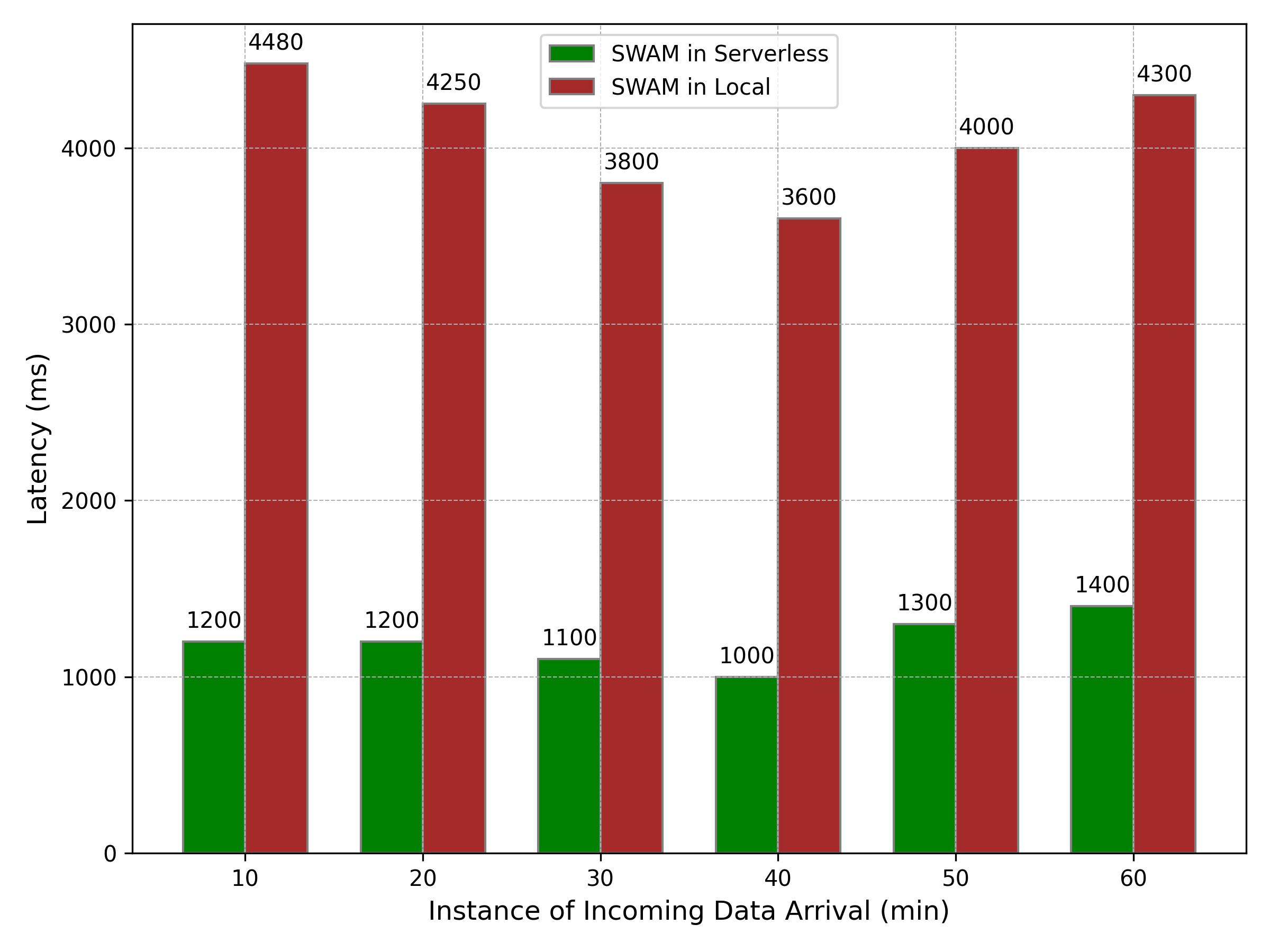}} \label{fig5a}
\subfigure[Avg. task waiting time (in ms)]{\includegraphics[width=0.45\textwidth,height=2in]{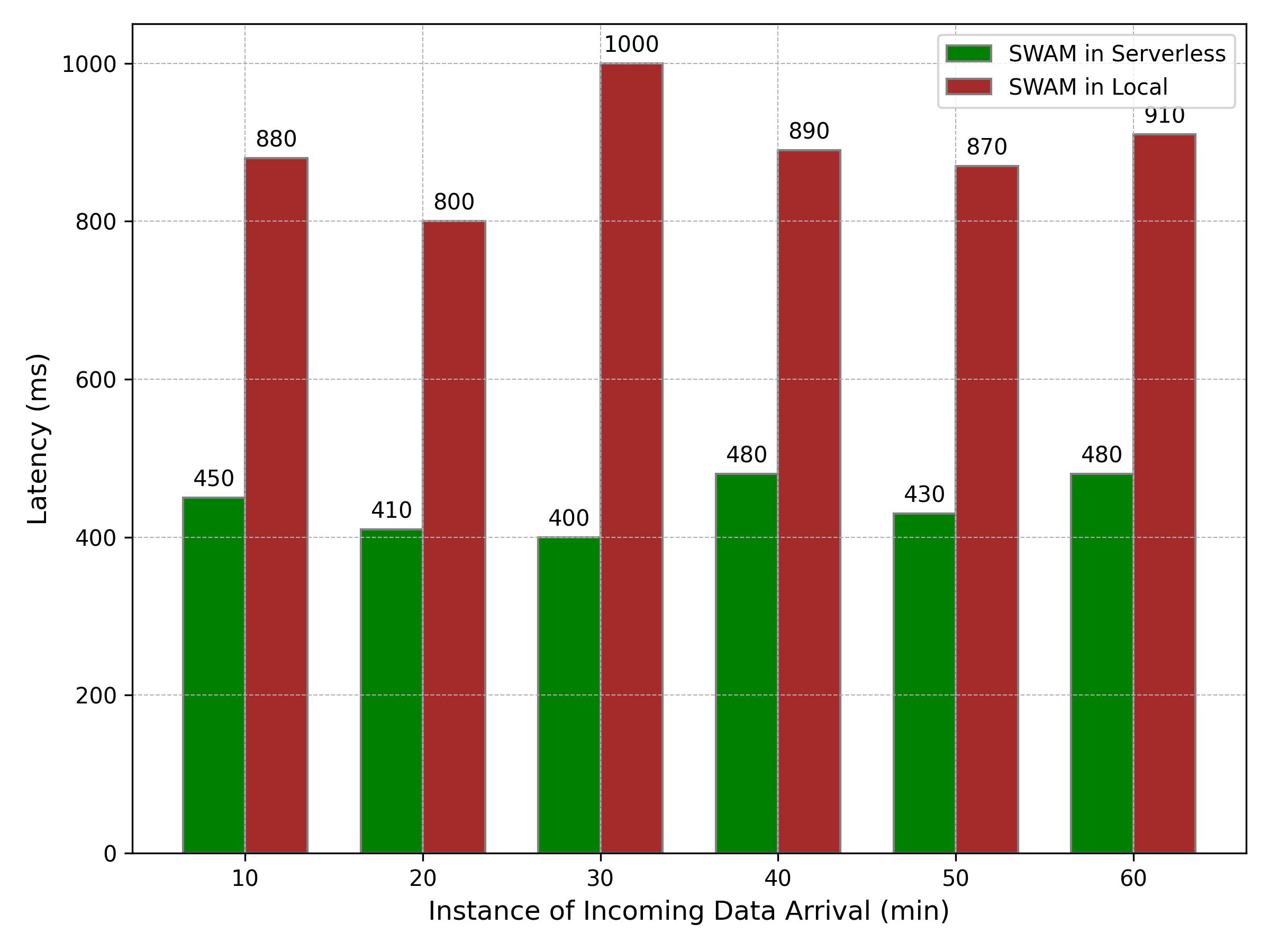}}  \label{fig5b}
\vspace{-0.1in}
\caption{Illustration of total end-to-end latency and average task waiting time of SWAM in serverless and local settings.}
\label{Fig5}
\vspace{-0.1in}
\end{figure}

\begin{figure}[!t]
\subfigure[Task completion ratio]{\includegraphics[width=0.45\textwidth,height=2in]{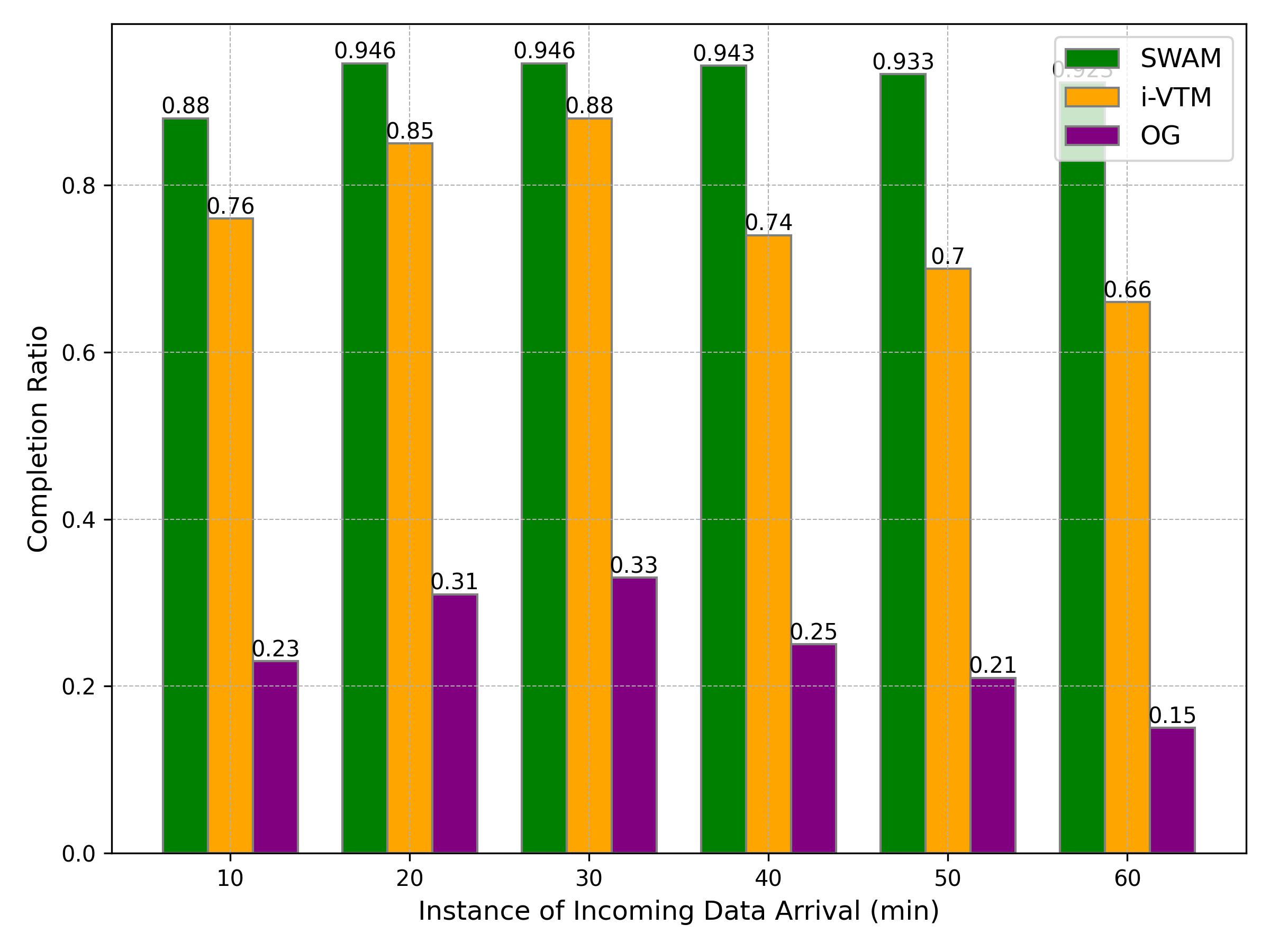}}\label{Fig6a}\hspace{-0.1in}
\subfigure[Overall utility score]{\includegraphics[width=0.45\textwidth,height=2in]{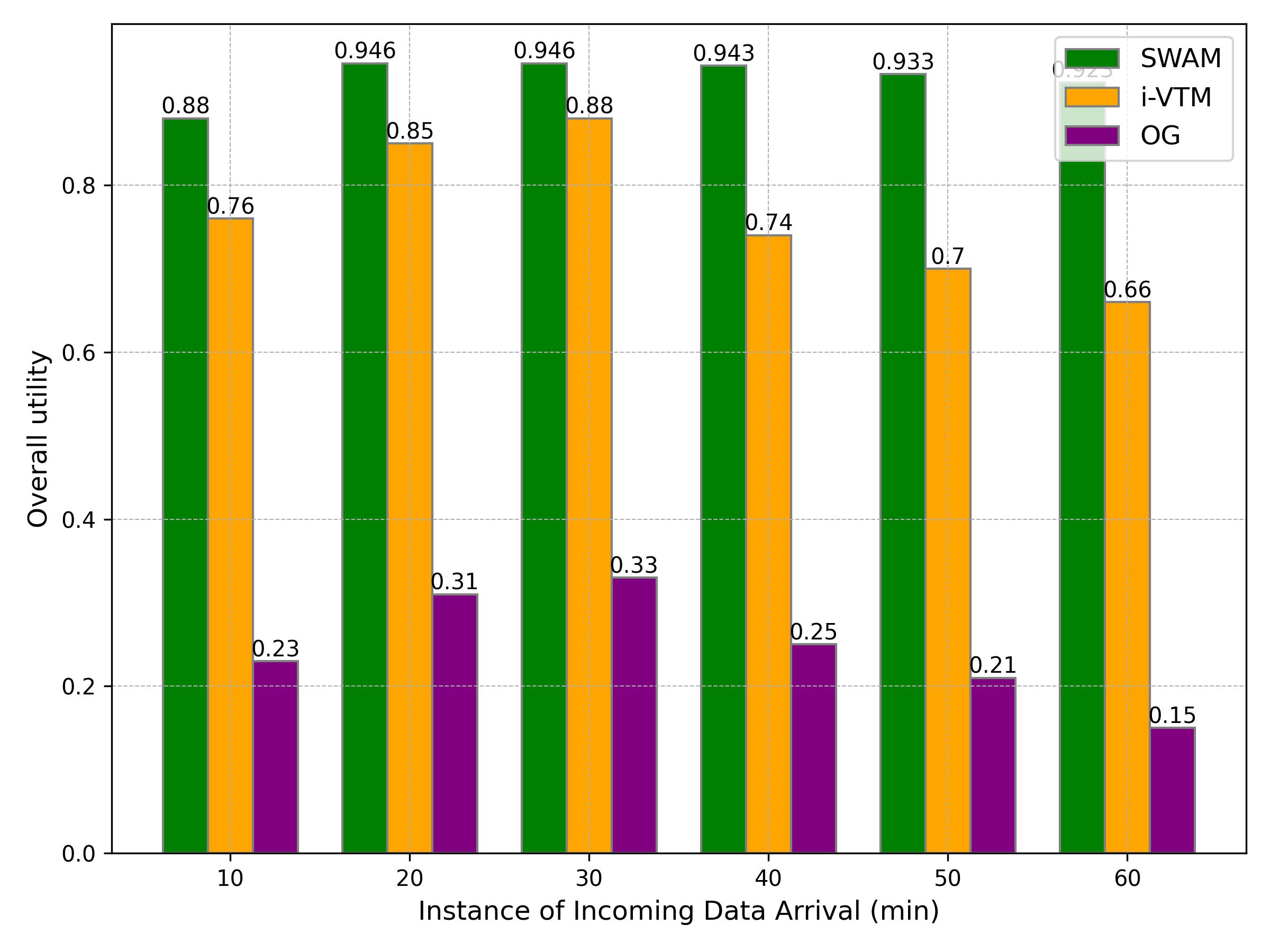}}\label{Fig6b}
\vspace{-0.1in}
\caption{Illustration of task completion ratio and overall utility score of SWAM, i-VTM and OG in the local setting.}
\label{Fig6}
\vspace{-0.1in}
\end{figure}

The previous analysis highlights the trend of the net utility score, depicted in \textbf{Figure 5(b)}. The average utility score for SWAM is 355, while i-VTM and OG achieve scores of 272 and 223, respectively. The inclusion of willingness as a factor further enhances SWAM's performance compared to i-VTM, underscoring its direct impact on utility scores. Additionally, the serverless computing framework significantly enhances the SWAM algorithm's performance. Leveraging Function-as-a-Service (FaaS), serverless computing enables parallel execution of data through modular, stateless functions. This modular execution accelerates data processing, and the platform's automatic scalability handles dynamic and unpredictable volunteer and task data influxes without developer intervention. The responsibility for configuring scalability resides entirely with the service provider, alleviating the developer's burden.

\vspace{-0.1in}
\section{Related Work}\label{relatedwork}
\subsection{Skill-based Task Assignment in Crowdsourcing}
Notwithstanding the advancements made, the present body of literature predominantly focuses on the allocation of micro-tasks that do not necessitate substantial cognitive exertion or specialized skills. This leaves a gap in the research regarding the allocation of tasks that require specific skills and cognitive effort. To fill this research gap, several studies have been initiated to explore the impact of skill orientation on task allocation in crowdsourcing. Nonetheless, within this body of literature, there is a scarcity of research that has explicitly concentrated on the allocation of tasks in volunteering services.

The article referenced as \cite{goel2014} addresses the challenge of connecting the skills of workers to the appropriate tasks in crowdsourcing markets. It proposes a mechanism that uses incentives to encourage workers with appropriate expertise to self-select tasks and the objective function includes the quality of work, the worker's expertise, and the cost of completing the task. Cheng et al. \cite{Cheng2016} propose an approach for task allocation in spatial crowdsourcing, where tasks demand multiple skills to be completed with two specific objectives. The first objective function minimizes the total time required to complete all tasks, including the time for task execution and worker travel time and the second objective function maximizes the total number of tasks completed within a specific time frame, while also considering the skills required for each task and the skills possessed by each worker. Liu et al. \cite{Liu2016} further advanced the problem of complex task allocation by proposing an approach that considers the tasks-workers-skills tripartite graph. The approach utilizes two objective functions, one which minimizes the total time required to complete all tasks and another which maximizes the total number of tasks completed within a given time frame. All the above studies are done in offline settings. The allocation of multi-skill tasks in online settings was studied in \cite{Song2020} and \cite{samanta2021swill}, with the latter adding a willingness component to the approach. Ni et al. \cite{Ni2020} defined dependency-aware spatial crowdsourcing, focusing on maximizing the number of successfully assigned tasks with constraints such as worker skills and deadlines. Liang et al. \cite{liang2022multi} put forward a cost-oriented greedy strategy for reducing platform expenses through the pairing of appropriate groups of workers. The selection of volunteers based on skills and proximity was discussed in \cite{Sama2212:Volunteer}. Riya et al. \cite{samantafogirecruiter} addresses a situation of choosing participants in crowdsourcing for disaster management.

\subsection{Serverless Computing}
A considerable amount of research has been done in the field of serverless computing. Papers such as \cite{liang2022multi} and \cite{hellerstein2018serverless} provides an overview of serverless computing, including its features, opportunities, and challenges. The scalability of serverless platforms and their potential impact on the industry have been discussed in \cite{9041780} and \cite{lee2018evaluation}. Additional research efforts, such as those presented in \cite{chiliman2021serverless} and \cite{zhang2019serverless}, investigate different aspects of serverless computing, including cost-effectiveness, performance, and security. Moreover, \cite{perez2019programming} propose a serverless architecture for highly parallel file-processing tasks and discuss a middleware implementation that facilitates the execution of custom runtime environments within serverless computing. Authors in \cite{barcelona2019faas} performed an analysis on various services of different commercially available serverless platforms which claim effective parallelism of workloads. \cite{bharti2021sequential} brings the concept of function orchestration and discusses different sequential composition workflows whereas \cite{lopez2020triggerflow} talks about an extensible Trigger-based Orchestration architecture for serverless workflows. The field of serverless computing is constantly evolving, and new research is being conducted to further enhance its capabilities and address existing limitations.

\section{Conclusion}\label{conc}
Matching tasks with the most suitable volunteers remains a significant challenge in volunteer crowdsourcing ventures. Complex tasks often necessitate collaboration among volunteers with diverse skill sets. In addition to skills, the willingness of volunteers to engage in community activities is crucial. These volunteer crowdsourcing (VCS) platforms regularly experience a dynamic influx of tasks and volunteers. Determining the required computing resources for volunteer to task matching process is challenging due to unpredictable platform demands in terms of volunteer and tasks requests Therefore, these platforms must strategically allocate tasks to volunteers in a scalable manner. The utilisation of serverless technology presents a feasible methodology.

In this paper, we proposed a serverless-assisted framework for task assignment in the VCS paradigm. The proposed Skill and Willingness-Aware Volunteer Matching (SWAM) algorithm allocates volunteers to tasks based on their skills, payments, willingness, and tasks’ skill requirements. We developed a serverless framework to deploy the SWAM algorithm, achieving a performance efficiency of 71\% in the serverless environment in terms of total end-to-end latency. The investigations were conducted using the real-world MeetUp dataset. We achieved a task completion ratio of 92\% and reduced the waiting time for submitted tasks before assignment to volunteers by 56\%. Furthermore, the overall utility gain was approximately 30\% higher than the baselines.

In future research, we plan to use cooperative game theory to model the collaborative dynamics among volunteers (workers), tasks, and platforms. Integrating cooperative game theory into the skill-oriented task assignment process will promote equitable, productive, and collaborative outcomes. Additionally, we aim to test our framework on a larger dataset and improve the algorithm for assigning tasks to volunteers. This will help in further refining the task matching process and enhancing the efficiency and effectiveness of volunteer crowdsourcing platforms. By continuing to innovate in this space, we hope to further support grassroots movements and foster greater citizen engagement and transparency, ultimately contributing to the global social good.

\bibliographystyle{ACM-Reference-Format}
\bibliography{sample-base}

\end{document}